\documentclass[sigconf,natbib=false]{acmart}
\usepackage{subfigure}
\usepackage{graphicx}
\usepackage{multirow}
\usepackage{booktabs}
\usepackage{diagbox}
\usepackage{amsmath}
\usepackage{algorithm}
\usepackage{algorithm}
\usepackage{algorithmicx}
\usepackage{algpseudocode}

\AtBeginDocument{%
  }

\copyrightyear{2023} 
\acmYear{2023} 
\setcopyright{acmlicensed}\acmConference[CIKM '23]{Proceedings of the 32nd
ACM International Conference on Information and Knowledge
Management}{October 21--25, 2023}{Birmingham, United Kingdom}
\acmBooktitle{Proceedings of the 32nd ACM International Conference on
Information and Knowledge Management (CIKM '23), October 21--25, 2023,
Birmingham, United Kingdom}
\acmPrice{15.00}
\acmDOI{10.1145/3583780.3615054}
\acmISBN{979-8-4007-0124-5/23/10}

\RequirePackage[
  datamodel=acmdatamodel,
  style=acmnumeric,
  ]{biblatex}

\begin{document}

\title{Self-Supervised Dynamic Hypergraph Recommendation based on Hyper-Relational Knowledge Graph}



\author{Yi Liu}
\orcid{0000-0001-7710-4578}
\affiliation{
  \institution{Nanjing University of Aeronautics and Astronautics}
  \city{Nanjing}
  \country{China}
  \postcode{211100}
}
\email{liuyi-sx21@nuaa.edu.cn}

\author{Hongrui Xuan}
\orcid{0000-0002-2435-2858}
\affiliation{
  \institution{Nanjing University of Aeronautics and Astronautics}
  \city{Nanjing}
  \country{China}
  \postcode{211100}
}
\email{1692595335@nuaa.edu.cn}

\author{Bohan Li}
\orcid{0000-0002-3408-9037}
\authornote{Corresponding author}
\affiliation{
  \institution{Nanjing University of Aeronautics and Astronautics}
  \city{Nanjing}
  \country{China}
  \postcode{211100}
}
\email{bhli@nuaa.edu.cn}

\author{Meng Wang}
\orcid{0000-0002-2293-1709}
\affiliation{
  \institution{Tongji University}
  \city{Shanghai}
  \country{China}
  \postcode{211100}
}
\email{wangmengsd@outlook.com}

\author{Tong Chen}
\orcid{0000-0001-7269-146X}
\affiliation{
  \institution{The University of Queensland}
  \city{Brisbane}
  \country{Australia}
}
\email{tong.chen@uq.edu.au}

\author{Hongzhi Yin}
\orcid{0000-0003-1395-261X}
\affiliation{
  \institution{The University of Queensland}
  \city{Brisbane}
  \country{Australia}
}
\email{h.yin1@uq.edu.au}


\begin{abstract}
Knowledge graphs (KGs) are commonly used as side information to enhance collaborative signals and improve recommendation quality. In the context of knowledge-aware recommendation (KGR), graph neural networks (GNNs) have emerged as promising solutions for modeling factual and semantic information in KGs. However, the long-tail distribution of entities leads to sparsity in supervision signals, which weakens the quality of item representation when utilizing KG enhancement. Additionally, the binary relation representation of KGs simplifies hyper-relational facts, making it challenging to model complex real-world information. Furthermore, the over-smoothing phenomenon results in indistinguishable representations and information loss.

To address these challenges, we propose the SDK (Self-Supervised Dynamic Hypergraph Recommendation based on Hyper-Relational Knowledge Graph) framework. This framework establishes a cross-view hypergraph self-supervised learning mechanism for KG enhancement. Specifically, we model hyper-relational facts in KGs to capture interdependencies between entities under complete semantic conditions. With the refined representation, a hypergraph is dynamically constructed to preserve features in the deep vector space, thereby alleviating the over-smoothing problem. Furthermore, we mine external supervision signals from both the global perspective of the hypergraph and the local perspective of collaborative filtering (CF) to guide the model prediction process.

Extensive experiments conducted on different datasets demonstrate the superiority of the SDK framework over state-of-the-art models. The results showcase its ability to alleviate the effects of over-smoothing and supervision signal sparsity.

\end{abstract}




\maketitle

\section{Introduction}

As an information retrieval tool, a recommender system searches for valuable information based on users' needs and interests, providing personalized recommendations ~\cite{gao2022graph}. In the Internet era, recommender systems have been successfully applied in various fields such as online video~\cite{wei2019mmgcn}, Internet advertising~\cite{sun2021fm2}, e-commerce~\cite{wang2018billion}, and more, achieving significant advancements and aiding decision-making processes by handling large amounts of data.

Collaborative filtering (CF) is a fundamental algorithm extensively used in recommender systems. Its core idea is to establish a preference model for users based on their historical interaction behaviors. For example, the matrix factorization method directly obtains users' and items' embeddings by decomposing the co-occurrence matrix, capturing their similarity in the vector space through inner product. However, the traditional CF methods fail to consider the high-order connection information. To address this, the graph neural network (GNN) framework like LightGCN~\cite{he2020lightgcn} is introduced to leverage multiple embedding propagation layers to capture collaborative signals in high-order connections. HCCF~\cite{xia2022hypergraph}, on the other hand, constructs a hypergraph structure for the user-item bipartite graph and integrates the encoding of global hypergraph structure and local collaborative effects to enhance the quality of representations. However, CF-based methods often focus solely on the user-item interaction matrix and neglect side information~\cite{wang2018tem, wang2017item}, leading to limitations in handling sparse cases with few connections. 

Recent studies have tried to overcome the limitations of CF-based methods by incorporating knowledge graphs (KGs) as side information to enhance the semantic information of items. KGs are graph-structured knowledge bases that store facts in the form of triplets, consisting of a head entity, a relation, and a tail entity, to describe the physical world~\cite{yan2020knowime}. In KG-based methods, items are often mapped to the KG, and item representations are obtained by leveraging relation extraction through neighborhood aggregation. For instance, KGAT~\cite{wang2019kgat} integrates the user-item bipartite graph with the KG to model high-order structures on a collaborative knowledge graph, thereby enhancing the collaborative signals between users and items and improving the prediction of user preferences. CKAN~\cite{wang2020ckan} introduces a heterogeneous propagation mechanism to determine the importance of knowledge-aware neighbors, and integrates the representation obtained by modeling the KG into the CF vector space.

While previous work has made significant progress in recommendation performance, there are still several challenges that have not been effectively addressed:

\begin{itemize}
  \item \textbf{Over-smoothing representation:} The phenomenon of over-smoothing refers to the convergence of embeddings in the vector space, where the model struggles to differentiate between vectors. As the network deepens, the message-passing framework in GNNs generates similar computation graphs for each node by aggregating features from neighboring nodes, causing the node representations of the same connected component to converge to the same value.  The dilemma is that higher-order information is essential to address the data sparsity issue in the recommendation system.
  \item \textbf{Supervision signal sparsity:} KG-based recommendation methods typically follow a supervised learning strategy, where user-item interactions are utilized to supervise the training process after enhancing the KG. The effectiveness of this way heavily relies on the availability of an adequate number of supervision signals. However, in practical recommendation applications, the collected interaction data is often limited and highly sparse, resulting in the issue of insufficient training labels for most users. The scarcity of supervision signals not only hinders the model from obtaining high-quality representations but also leads to degradation problems.
\end{itemize}

\begin{figure}[htbp]
  \centering
  \includegraphics[width=\linewidth]{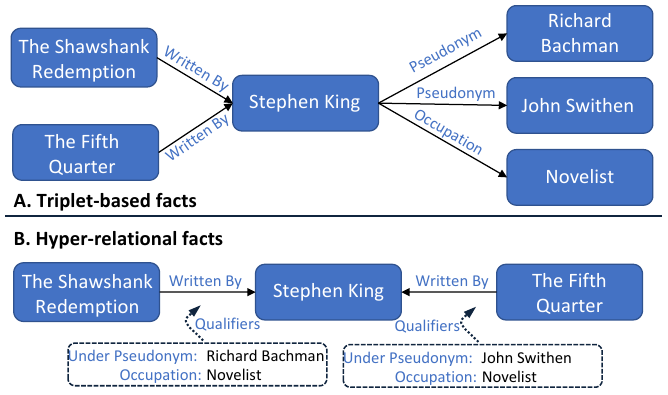}
  \caption{Example of triplet-based facts and hyper-relational facts.}
  \label{intro}
\end{figure}

Given the aforementioned challenges, we propose the "Self-Supervised Dynamic Hypergraph Recommendation based on Hyper-Relational Knowledge Graph" (SDK) framework to improve the representation ability and generalization performance of KG-based recommendation paradigms. Specifically, we transform the KG into a hyper-relational format and model facts within the context of $N$-ary relations. This approach enables us to capture more nuanced relationships among entities. For example, a hyper-relational fact is depicted in Figure 1. In the expression of the triplet-based facts $(h,r,t)$, we can know that `The Fifth Quarter' and `The Shawshank Redemption' are both written by Stephen King. Additionally, the entity set indicates that Stephen King is a novelist with the pen names Richard Bachman and John Swithen. However, it becomes confusing as we cannot determine which book was written under which pen name. The form of hyper-relational facts, which consists of a basic triplet $(h, r, t)$ and several qualifiers $(qr, qv)$ (Under Pseudonym, Richard
Bachman), allows us to observe that Stephen King wrote these two books with different pen names. Unlike triplet-based facts that model each piece of semantic information independently before aggregation, hyper-relational facts represent intrinsic semantic associations by directly modeling the basic triplet and qualifiers as a whole. By employing the SDK framework, we aim to enhance the representation ability and generalization performance of KG-based recommendation systems by leveraging hyper-relational facts and their qualifiers. This approach enables us to capture more detailed and nuanced relationships among entities, thereby improving the accuracy and effectiveness of recommendations. Additionally, the self-supervised nature of the framework ensures that it can adapt and learn from the data without requiring explicit annotations or labels, making it more scalable and applicable in real-world scenarios.

Furthermore, we construct a hypergraph based on hyper-relational semantic representation and adopt the hypergraph learning framework to model global dependencies, which serves as a supplement to encoding high-order relations. This approach is motivated by the understanding that initial semantics may not be suitable for data representation. To address this, we utilize adjusted feature embedding to dynamically and adaptively modify the hypergraph, thereby enhancing the robustness of its structure. While our SDK framework draws inspiration from the works~\cite{wu2021self,yang2022knowledge}, who utilize self-supervised data augmentation methods to provide additional supervision signals, there is a key difference in our approach. Instead of enhancing interaction data by perturbing graph structures, we combine local collaborative relations and global hypergraph dependencies to distill supervision signals. These signals then guide our SDK framework to refine user/item representations. By incorporating hyper-relational semantic representation, hypergraph learning, and adjusted feature embedding, our SDK framework aims to capture the intricate relationships among entities and improve the representation ability and generalization performance of KG-based recommendation paradigms. The combination of local collaborative relations and global hypergraph dependencies enables us to leverage rich semantic information while maintaining the structural integrity of the hypergraph.

Overall, our contributions are summarized as follows:
\begin{itemize}
\item Under the self-supervised learning paradigm, we integrate KG with user-item interactions and apply hypergraph learning to improve the robustness of recommendations, as well as alleviate the problems of data sparsity and over-smoothing.
\item We model the hyper-relational KG (HKG), mining the connections of instances in the graph under $N$-ary semantics, which enhances the semantic richness of the item vector space.
\item To collaboratively supervise the instance discrimination process and guide the model in capturing intrinsic relations, we integrate both local and global learning views.
\item Extensive experiments demonstrate that the SDK framework achieves significant improvements compared to state-of-the-art recommendation methods. Additionally, relevant ablation studies have validated the rationale behind each module.
\end{itemize}

\section{Related Work}

\textbf{KG-based Recommendation:} Introducing KG as side information to enrich item connection edges enables more accurate and interpretable personalized recommendations.
There are usually three types: (1) embedding-based methods~\cite{zhang2016collaborative,wang2018dkn} directly applied graph embedding methods to mine the semantic information between entities and relations in KG as context to enrich the representation of entities in the recommender system.
For example, CKE~\cite{zhang2016collaborative} encodes the structured knowledge of the entities through TransR~\cite{lin2015learning}, combines the context knowledge of text and images to obtain the semantic features, and fuses them into vector space.
DKN~\cite{wang2018dkn} applies TransD~\cite{ji2015knowledge} to model entities in news articles and obtain high-order semantic information about news.
(2) path-based methods~\cite{hu2018leveraging,shi2018heterogeneous,wang2019explainable,chen2021temporal} enhance the representation by constructing meta-paths in advance or automatically to mine the connection patterns between entities, and have stronger interpretability.
RKGE~\cite{sun2018recurrent} feeds paths between users and items into Recurrent Neural Network (RNN) as a sequence to capture connection patterns and automatically learn semantic representations.
HeteRec~\cite{yu2014personalized} obtains the corresponding user preference matrix from a variety of different types of meta-paths and combines the user's preferences on each path to generate recommendations.
(3) GNN-based methods~\cite{wang2019knowledge, wang2020ckan,xuan2023knowledge,zou2022improving} leverage the message propagation and aggregation mechanism of GNN to capture multi-hop neighbor information and dependencies between nodes in an iterative manner.
For example, KGAT~\cite{wang2019kgat} integrates user-item bipartite graph and KG, and iteratively aggregates neighbors to update embedding in the form of heterogeneous graph.
MVIN~\cite{zou2022multi} aggregates neighbor information in a user-oriented manner by considering user-entity interactions.

\noindent \textbf{Hyper-relational KG:}
Since the triplets in the traditional KG oversimplify the complexity of the data, recent studies have begun to model hyper-relational facts.
m-TransH~\cite{wen2016representation} is a method based on TransH~\cite{wang2014knowledge} to transform hyper-relational facts through star-to-clique conversion.
RAE~\cite{zhang2018scalable} builds upon m-TransH and further transforms hyper-relational facts into $N$-ary facts with abstract relations.
NaLP~\cite{guan2019link} proposes a link prediction method that models $N$-ary facts as role-value pairs and utilizes a convolution-based framework to compute the similarity of each pair.
StarE~\cite{galkin2020message} specifically designed an encoder for $N$-ary facts to be compatible with indefinite-length qualifiers and emphasize the interaction of the basic triplets to qualifiers.
QUAD~\cite{shomer2022learning} takes into account the bidirectional flow between base triplets and qualifiers by constructing a basic aggregator and a qualifier aggregator, respectively, which enhances the flow of the qualifier to the basic triplet while maintaining the valid impact of the qualifier on the basic triplet.
In general, the modeling of $N$-ary facts is usually applied to link prediction scenarios, but reasonable modeling of hyper-relational facts before link prediction to obtain entity representation is similar to traditional KG modeling.

\noindent \textbf{Self-supervised Hypergraph Learning for Recommendation:}
Inspired by the expressiveness of hypergraphs in complex high-order dependencies, several recent recommender systems introduce hypergraph neural networks to improve relation learning~\cite{yu2021selfsuper,xia2021self}.
Since the hypergraph view is derived from interaction data rather than perturbing the graph structure, self-supervised learning can guide the model from a robust perspective to improve embedding quality~\cite{qiu2022contrastive, yu2023xsimgcl,yu2022are, Yu2023self,}.
For example, SBR~\cite{xia2021self} proposes a dual-channel hypergraph convolutional network to capture cross-session information and innovatively integrates self-supervised learning tasks to maximize mutual information.
HCCF~\cite{xia2022hypergraph} constructs the global self-augmented contrastive view in the form of hypergraph, and collaboratively captures implicit dependencies based on the traditional CF paradigm.
SHT~\cite{xia2022self} applies a topology-aware Transformer in the framework of hypergraph neural network to realize the encoder of global collaborative effect, and combines with local view to realize graph topology denoising and knowledge distillation.

\section{Preliminaries}
\textbf{User-item Bipartite Graph:} In recommender systems, historical user-item interactions are usually represented as bipartite graphs.
Let $U=\{u_1, u_2, \cdots, u_N \}$ and $V=\{v_1, v_2, \cdots, v_M \}$ denote the user set and item set, respectively, where $N$ and $M$ represent the number of users and items.
The user-item bipartite graph is then $\mathcal{G}_u=\{y_{uv} \vert u \in U, v \in V \}$, where $y_{uv}=1$ if there is an observed interaction between user $u$ and item $v$, and $y_{uv} = 0$ if not.

\noindent \textbf{Knowledge Graph:} A KG provides auxiliary information for the recommender system to alleviate the problem of data sparsity. The KG $\mathcal{G}_k$ utilizes the triplet set $\{(h,r,t) \vert h,t \in E, r \in R \}$ to describe facts, where $E$ and $R$ are respectively the sets of entities and relations, and $(h,r,t)$ indicates there is a relation $r$ from head entity $h$ to tail entity $t$.
In KG-based recommendation where $V\in E$, an item $v \in V$ may form of triplets with several different entities
in the given KG $\mathcal{G}_k$.

\noindent \textbf{Hyper-relational Knowledge Graph:}
HKG is an extension of the standard KG, which describes the $N$-ary facts in the real world by supplementing the basic triple semantics with qualifier pairs.
A hyper-relational fact is represented by a tuple $(h, r, t, \mathcal{Q}_{hrt})$, where $(h, r, t)$ is the knowledge triplet, and $\mathcal{Q}_{hrt}$ is the set of qualifier pairs $\{(qv_i, qr_i)\}_{i=1}^{|\mathcal{Q}_{hrt}|}$ with qualifier relations $qr \in R$ and qualifier entities $qv \in E$.
In this way, these entities can be seen as connected by a hyperedge in the HKG, representing an $N$-ary fact, also known as a statement.

\noindent \textbf{Task Formulation:} Based on the given input $\mathcal{G}_u$ and $\mathcal{G}_k$, our purpose is to learn a function to predict whether the user $u$ is interested in an item $v$ that has not been interacted before, defined as $\hat{y}_{uv}=\mathcal{F}(u,v, \Theta, \mathcal{G}_u,\mathcal{G}_k)$, where $\Theta$ represents the model parameters and $\hat{y}_{uv}$ represents the predicted probability that user $u$ adopts the target item $v$.

\section{Methodology}

\begin{figure*}[!t]
  \centering
  \includegraphics[angle=0,width=1\textwidth]{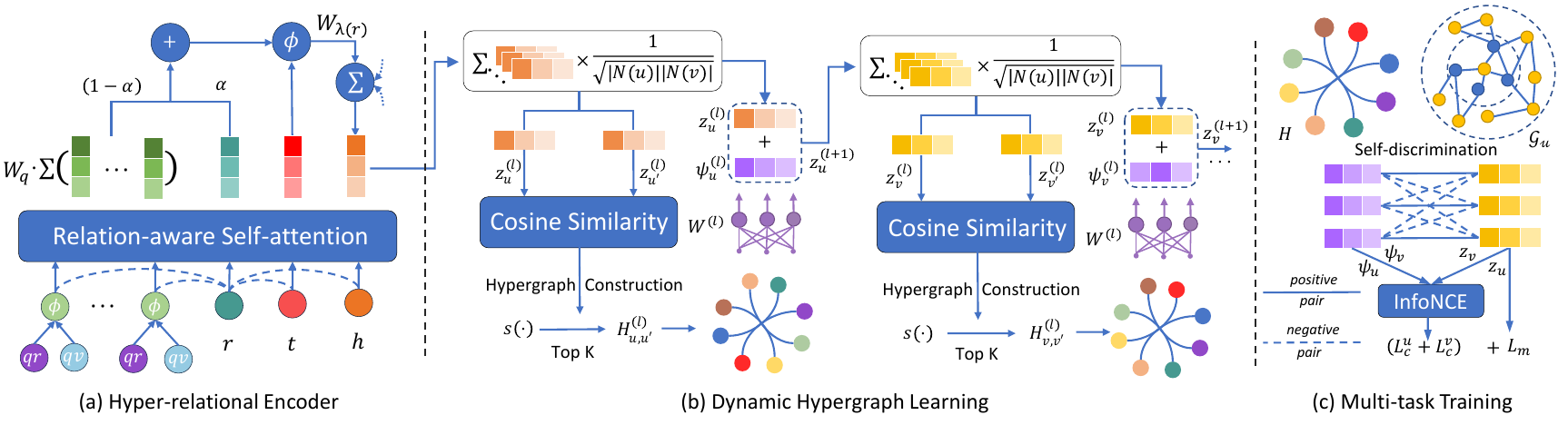}
  \caption{The overview of the HKG enhanced architecture based on dynamic hypergraph learning. (a) Capturing intrinsic semantic associations in hyper-relational facts and aggregating into $x_h$. (b) Dynamically construct hypergraph and capture global dependencies $\boldsymbol{\psi}^{(l)}$ by propagation. (c) A joint learning paradigm to generate contrastive losses $L_{c}^u$ and $L_{c}^i$.}
  \label{overflow}
\end{figure*}

The overall workflow of our proposed SDK is shown in Figure \ref{overflow}.
In the following sections, we describe the following key components of SDK: (1) Hyper-relational Encoder which captures intrinsic dependencies under $N$-ary semantics;
(2) Dynamic Hypergraph Learning which constructs hypergraphs adaptively and models global dependencies; and
(3) Multi-task Learning that combines local and global perspectives for joint model optimization.

\subsection{Hyper-relational Encoder}
\subsubsection{Relation-aware Self-attention}
The purpose of the hyper-relational encoder is to extract and leverage the rich semantics within the HKG. To fully exploit the $N$-ary knowledge in hyper-relational facts, we start by capturing the interdependencies between entities with the observed, complete semantics.
To achieve this, we first integrate each qualifier pair as follows:
\begin{equation}
  \boldsymbol{e}_q = \phi(\boldsymbol{e}_{qv},\boldsymbol{e}_{qr})
\end{equation}
\noindent where $\boldsymbol{e}_{qv}$ and $\boldsymbol{e}_{qr}$ are respectively entity and relation embeddings in a qualifier pair $\mathcal{Q}_{hrt}$.
$\phi(\cdot)$ is an interaction function between embeddings, which can be subtraction, multiplication, and rotation operations.
Next, by collecting the initial embeddings of the basic triplet $(h, r, t)$, denoted by $\boldsymbol{e}_h$, $\boldsymbol{e}_r$, and $\boldsymbol{e}_t$, and all the integrated qualifier representations $\boldsymbol{e}_q$ for $q \in\mathcal{Q}_{hrt}$, we can result in a set of $S$ input embeddings $\{\boldsymbol{e}_1, \boldsymbol{e}_2, \cdots, \boldsymbol{e}_S\} \in \mathbb{R}^d$ ($S=3+|\mathcal{Q}_{hrt}|$). Note that we simplify the subscripts into $1,2,\cdots,S$ when there is no ambiguity.
With the input embeddings, we obtain the updated representations $(\boldsymbol{x}_1, \boldsymbol{x}_2, \cdots, \boldsymbol{x}_S) \in \mathbb{R}^d$ by mining the correlations among inputs via self-attention~\cite{Vaswani2017attention}:
\begin{equation}
  \begin{split}
    \pi_{ij} &= \frac{(W_Q \boldsymbol{e}_i)^T (W_K \boldsymbol{e}_j + \boldsymbol{a}_{ij}^K) }{\sqrt{d}} \\
    \tilde{\pi}_{ij} &= softmax(\pi_{ij}) = \frac{exp{(\pi_{ij})}}{\sum_{j\prime =1}^{\vert S \vert}exp{(\pi_{ij^\prime})}} \\
    \boldsymbol{x}_i &= \sum_{j=1}^{S} \tilde \pi_{ij}(W_V \boldsymbol{e}_j + \boldsymbol{a}_{ij}^V)
  \end{split}
\end{equation}
\noindent where learnable weights $W_Q,W_K,W_V \in \mathbb{R}^{d \times d}$ are adopted to transform the input into the same vector space.
For the HKG, entities $h, t \in E$ and corresponding qualifier pairs $(qv, qr)\in \mathcal{Q}_{hrt}$ are highly dependent on relation $r$, hence it is necessary to consider the attentive bias of $r$ to other instances in detail.
Inspired by~\cite{shaw2018self}, we introduce learnable biases $\boldsymbol{a}_{ij}^K$, $\boldsymbol{a}_{ij}^V \in \mathbb{R}^d$ to obtain edge-aware attention between $i$ and $j$.
In particular, we explicitly encode the interactions between $r$ and other instances, which are used for supplementing the attentive bias to be learned.
If there is no interaction between $i$ and $j$, then we set $\boldsymbol{a}_{ij}^K=\boldsymbol{0}$, $\boldsymbol{a}_{ij}^V=\boldsymbol{0}$.
Unlike the traditional self-attention mechanism that only encodes the global dependency between two embeddings, the introduction of $\boldsymbol{a}_{ij}^K$, $\boldsymbol{a}_{ij}^V$ captures additional local dependency of the instance on $r$. This allows us to fully account for the intricate correlations between qualifier pairs and relations, providing a more comprehensive semantic context for the downstream recommendation.

\subsubsection{Hyper-relational Aggregator}
With an analogous structure to traditional KGs, an HKG can take advantage of similar approaches for updating entity and relation representations.
In this regard, a typical work for modeling HKG is StarE~\cite{galkin2020message}, which bears a GNN-based architecture as follows:
\begin{equation}\label{eq:HRA}
  \boldsymbol{z}_h = f\bigg(\sum_{(r,t)\in N(h)}W_{\lambda(r)} \phi(\boldsymbol{x}_t, \gamma(\boldsymbol{x}_r, \boldsymbol{x}_{qs}))\bigg)
\end{equation}
\noindent where $\phi(\cdot)$ is the interaction function, and $N(h)$ represents the triplet associated with head $h$. $\boldsymbol{x}_{qs} \in \mathbb{R}^{d}$ is the merged embedding of qualifiers, $\gamma(\cdot)$ is the weighted sum function, both of which will be introduced in detail. $W_{\lambda(r)}$ is a projection matrix specific to relation $r$'s direction $\lambda(r) =\{\text{forward, reverse, self-loop}\}$, and $f(\cdot)$ is a nonlinear activation function. However, this approach only aggregates neighbor nodes around $h$ without considering the influence of head entity $h$ itself. To fill this gap, our hyper-relational aggregator captures $N$-ary information additionally integrates the semantics of $h$, where we propose a generalized approach for modeling hyper-relational information:
\begin{equation}
  \boldsymbol{z}_{h} = f\bigg(\frac{1}{\vert N(h)\vert }\sum_{n=1}^{\vert N(h)\vert} {\boldsymbol{x}_{h_n}} + \sum_{(r,t)\in N(h)}W_{\lambda(r)} \phi(\boldsymbol{x}_t, \gamma(\boldsymbol{x}_r, \boldsymbol{x}_{qs}))\bigg)
\end{equation}
\noindent where $\boldsymbol{x}_{h_n}$ is the representation of $h$ in each $N$-ary statement. Following the common practice in KG-based recommendation, we only consider $r$ where $\lambda(r)=\text{forward}$.
For qualifiers, we integrate all qualifier pairs and project them into the vector space of $r$ with weight matrix $W_q\in \mathbb{R}^{d \times d}$, and then obtain $\boldsymbol{x}_{qs}$ through the weighted sum function $\gamma(\cdot)$ as follows:
\begin{equation}
  \begin{split}
    \boldsymbol{x}_{qs} & =  W_q \sum_{q \in \mathcal{Q}_{hrt}} \boldsymbol{x}_q                         \\
    \gamma(x_r, x_{qs}) & = \alpha \odot \boldsymbol{x}_r + (1- \alpha) \odot \boldsymbol{x}_{qs}
  \end{split}
\end{equation}
\noindent where $\alpha$ is a hyperparameter to balance the contributions from $\boldsymbol{x}_r$ and $\boldsymbol{x}_{qs}$.

\subsection{Dynamic Hypergraph Learning}
In this section, we propose our dynamic hypergraph learning method, which captures local dependencies through message passing, dynamically constructs hypergraph structures based on the obtained refined representations, and then models global dependencies through hypergraph convolution layers.

\subsubsection{Dependency Hypergraph Construction}
For local dependencies, we first consider the collaborative signals between users and items within the latent space.
The well-established LightGCN~\cite{he2020lightgcn} is adopted as the base recommendation model for its simplicity and efficiency.
Taking users $u\in U$ as an example, we can obtain the following with $\mathcal{G}_u$:
\begin{equation}
  \begin{split}
    \boldsymbol{z}_u^{(l)} &= \sum_{v \in N(u)} \frac{\boldsymbol{z}_v^{{(l)}}}{\sqrt{\vert N(u) \vert \vert N(v) \vert}} \\
    c_{uu'}^{(l)} &= s(\boldsymbol{z}_u^{(l)}, \boldsymbol{z}_{u'}^{(l)})
  \end{split}
\end{equation}
\noindent where $\boldsymbol{z}_u^{(l)}$ denotes a user embedding at layer $l$, while $N(u)$ and $N(v)$ are respectively the neighbor set for user $u$ and item $v$. Note that for the first layer $l=1$, each item embedding $\boldsymbol{z}_v^{(l)}$ used for aggregation is the knowledge representation generated by Eq.(\ref{eq:HRA}). $c_{ij}$ quantifies the dependency between users $u$ and $u$, obtained by the cosine similarity function $s(\cdot)$.
A greater $c_{ij}$ indicates a higher dependency between users.
For each user $u$, we select the top $K$ relevant users $R(u)$, and connect them via a hyperedge to construct a user-user hypergrah $H^{(l)} \in \mathbb{R}^{\vert U \vert \times \vert U \vert}$:
\begin{align}
  H^{(l)}_{u,u'} =
  \begin{cases}
    c_{uu'}^{(l)} & \,\, u' \in R(u), \forall u \in U \\
    0       & otherwise
  \end{cases}
\end{align} 
\noindent where each row $H^{(l)}_{u}$ is a hyperedge marking a set of $K$ most relevant users $u'$ to $u$, and non-zero entries $H^{(l)}_{u,u'} = c_{uu'}^{(l)}$ quantifies the importance of $u'$ within the hyperedge. Analogously, the item-item hypergraph $H^{\prime(l)}\in \mathbb{R}^{\vert I \vert \times \vert I \vert}$ can also be constructed by mining the similarity between item embeddings. Since the low-level embeddings has limited expressiveness for capturing high-order dependencies, the intrinsic structure of the resulting hypergraph tend to be less stable and may weaken the learned hyperedge representations.
In this regard, we perform the above process for each layer $l$, so that the composition of hyperedges is adaptive to layer-specific user/item embeddings and will in turn gradually refine embeddings in the subsequent layer.

\subsubsection{Dynamic Hypergraph Convolution}
\noindent To capture the global high-order dependencies based on hyperedges, we follow the hypergraph message passing paradigm to construct a hypergraph convolutional layer. Still, with user-user hypergraph $H^{(l)}$ as an example, all users' hyperedges are learned through the following: 
\begin{equation}
  \begin{split}
    \tilde H &= D_v^{-\frac{1}{2}}H^{(l)}D_e^{-\frac{1}{2}} \\
    \varPsi^{(l)} &= \sigma(\tilde H \tilde H^T Z^{(l)} W^{(l)})
  \end{split}
\end{equation}
\noindent where $\sigma(\cdot)$ is the LeakyReLU activation function, $D_v, D_e \in \mathbb{R}^{\vert U \vert \times \vert U \vert}$ are diagonal matrices representing the degree of nodes and hyperedges, respectively.
We obtain $\tilde H$ by normalization, which avoids numerical instability and gradient vanishing caused by stacking multiple hypergraph convolution layers. The second step is to aggregate the corresponding hyperedge representations according to $\tilde H$ to obtain all users' hyperedge representations $\varPsi^{(l)}\in \mathbb{R}^{|U|\times d}$, where $Z^{(l)} \in \mathbb{R}^{|U|\times d}$ stacks all $|U|$ user embeddings from the $l$-th layer of LightGCN, and $W^{(l)} \in \mathbb{R}^{d\times d}$ is the weight to be learned. 

After convolving on both the user-item interaction graph and the hypergraph, for each user $u$ we acquire two types of representations, namely $\boldsymbol{z}_u^{(l)}$ with local dependencies and $\boldsymbol{\psi}_u^{(l)}$ with global dependencies. Then, at the $(l+1)$-th layer, the user representation leverages information from both sides:
\begin{equation}
  \boldsymbol{z}_u^{(l+1)} = \boldsymbol{\psi}_u^{(l)} + \boldsymbol{z}_u^{(l)}
\end{equation}
\noindent 
where $\boldsymbol{z}_u^{(l+1)}$ is then used for graph convolutions at layer $l+1$. Notably, by propagating the users' features to items, the above process can be similarly applied to obtain updated item embeddings $\boldsymbol{z}_v^{(l+1)}$.
In SDK, we iteratively update the representations of all users and items by alternating their dynamic hypergraph learning processes.

\subsection{Multi-task Training}
The final step is to establish an objective function to optimize the model parameters of SDK.
For the recommendation task, we adopt the pairwise marginal loss $L_m$ as follows:
\begin{equation}
  L_m = \sum_{u \in U}\sum_{v\in N(u)} \sum_{v^\prime \notin N(u)} max(0, 1-\hat{y}_{uv} +\hat{y}_{uv^\prime})
\end{equation}
\noindent where $N(u)$ all items visited by user $u$, and $v^\prime$ represents the negative samples with no observed interactions.
$\hat{y}_{uv}$ denotes the matching score of user $u$ and item $v$ calculated by inner product $\hat{y}_{uv}=\boldsymbol{z}_u^{*T} \boldsymbol{z}_v^*$, where $\boldsymbol{z}_u^{*}$ and $\boldsymbol{z}_v^{*}$ are the final user/item representations by performing average pooling \cite{he2020lightgcn} of each user/item embeddings at all convolution layers.

Because the user-item interaction graph and the hypergraph respectively encode local collaborative information and global dependencies between users and items, to make full user of the auxiliary supervision signals, we construct a self-supervised learning scheme. Specifically, we contrast different views of a user/item instance by drawing learned representations from both the local and global contexts:
\begin{equation}
  L_{c}^u = \sum_{u \in U} -log \frac{exp(s(\boldsymbol{z}_u, \boldsymbol{\psi}_u) / \tau)} {\sum_{u^\prime \in U, u \neq u^\prime } exp(s(\boldsymbol{z}_u, \boldsymbol{\psi}_{u^\prime}) / \tau)}
\end{equation}
\noindent where $\tau$ is the temperature parameter adopted to control the smoothness of the softmax curve. Essentially, the self-supervision is facilitated by InfoNCE~\cite{chen2020simple} to maximize agreement/disagreement between the same/different instances from two views.
$L_c^u$ represents the contrastive loss on the user side, while we analogously obtain the contrastive loss on the item side $L_c^v$.
Finally, we integrate hypergraph contrastive loss into the recommendation loss as a multi-task learning process as follows:
\begin{equation}
  L = L_{m} + \lambda_1 (L_{c}^u + L_{c}^v) + \lambda_2\|\Theta\|^2_2
\end{equation}
\noindent where $\lambda_1$ is adopted to control the balance between two loss terms, and $\lambda_2$ is the hyperparameter that weighs the $L_2$ regularization term posed on all trainable parameters $\Theta$.

\section{Experiments}
Through extensive experiments, we demonstrate the effectiveness of SDK in this section by answering the following research questions (RQs):
\begin{itemize}
  \item \textbf{RQ1:} How does our proposed SDK perform compared with the state-of-the-art recommendation models?
  \item \textbf{RQ2:} How do the proposed key components in our SDK contribute to the overall performance?
  \item \textbf{RQ3:} Can SDK effectively alleviate the problem of over-smoothing and supervision signal sparsity?
  \item \textbf{RQ4:} What are the effects of different hyperparameters on the performance of SDK framework?
\end{itemize}

\subsection{Experimental settings}
\subsubsection{Datasets}
To evaluate the performance of SDK, we conduct experiments on three open datasets collected from real-world applications and frequently used in recommendation tasks, \textit{i.e.}, Yelp2018, Amazon-book and MIND.
We follow \cite{galkin2020message} and construct the HKG for Yelp2018, Amazon-book, and MIND by extracting entities with main object and qualifier values in Wikidata.
Table \ref{table1} shows the statistics of all datasets, and their descriptions are as follows:
\begin{itemize}
  \item \textbf{Yelp2018} is a dataset of business venues from the 2018 Yelp Challenge, containing about one million interactions on 45,919 items from 45,538 users.
  \item \textbf{Amazon-book} is a dataset of book reviews collected on Amazon's e-commerce platform and commonly used for KG-based recommendations.
        It contains about one million interactions on 24,915 items from 70679 users.
  \item \textbf{MIND} is a news recommendation dataset constructed by the user click log of Microsoft News, containing about two million interactions on 48,957 items from 63,741 users. 
\end{itemize}

\begin{table}[h!]
  \begin{center}
    \caption{Statistics of datasets.}
    \label{table1}
    \begin{tabular}{c|c|c|c}
      \hline
      \rule{0pt}{10pt}
                       & \textbf{Yelp2018} & \textbf{Amazon-book} & \textbf{MIND} \\
      \hline
      \# Users          & 45,919            & 70,679               & 63,741        \\
      \# Items          & 45,538            & 24,915               & 48,957        \\
      \# Interactions   & 1,183,610         & 846,434              & 1,874,253     \\
      \hline
      \# Relations      & 51                & 43                   & 86            \\
      \# Entities       & 47,472            & 29,714               & 92,763        \\
      \# Statements     & 936,278           & 621,713              & 683,965       \\
      \# w / Quals (\%) & 433,496 (46.3)    & 256,145 (41.2)       & 84,073 (12.3) \\
      \hline
    \end{tabular}
  \end{center}
\end{table}

\subsubsection{Baselines}
We demonstrated the advantage of our SDK by comparing it to the recommendation models in different categories. The details of these models are as follows:

\noindent \textbf{Collaborative Filtering Methods}
\begin{itemize}
  \item \textbf{BPR}~\cite{rendle2014bayesian} performs Bayesian analysis from implicit feedback to obtain the maximum a posteriori probability to generate recommendation rankings.
  \item \textbf{NCF}~\cite{he2017neural} employs multi-layer perceptrons to achieve linear or nonlinear modeling of user-item interactions to obtain hidden features.
\end{itemize}
\noindent \textbf{GNN-based Methods}
\begin{itemize}
  \item \textbf{KGAT}~\cite{wang2019kgat} integrates KG and user-item bipartite graphs, and initializes feature embedding through transR.
        In addition, it adopts the attention mechanism to control the weight of multi-hop neighbors in the propagation process to achieve differentiated aggregation of information.
  \item \textbf{LightGCN}~\cite{he2020lightgcn} follows the basic GNN paradigm, but removes the nonlinear modules to achieve lightweight design and efficient training.
  \item \textbf{MVIN}~\cite{zou2022multi} proposes that differentiated user perspectives play a guiding role in neighborhood aggregation, and then generate its unique item view for each user, effectively modeling users' unique preferences.
  \item \textbf{CKAN}~\cite{wang2020ckan} utilizes heterogeneous propagation strategies to explicitly encode collaborative signals and knowledge associations to enrich node embedding, and applies knowledge-aware attention mechanisms to allocate weights during propagation.
\end{itemize}
\noindent \textbf{Hypergraph Neural Networks for Recommendation}
\begin{itemize}
  \item \textbf{DHCF}~\cite{ji2020dual} proposes a dual channel hypergraph CF framework to learn the representation of users and items in a divide and conquer way, and then uses hypergraph structure to model high-order dependencies.
  \item \textbf{HCCF}~\cite{xia2022hypergraph} establishes a hypergraph structure on the user-item interaction graph and captures the global collaborative relations. It adopts self-supervised learning to combine local and global perspectives in a cross-view manner to enhance the discrimination ability of the CF paradigm.
\end{itemize}
\noindent \textbf{Self-Supervised Learning for Recommendation}
\begin{itemize}
  \item \textbf{SGL}~\cite{wu2021self} applies the graph enhancement method to perturb the graph structure from multiple dimensions to generate different views and maximize the similarity of different views of the same instance.
  \item \textbf{KGCL}~\cite{yang2022knowledge} provides additional supervision signals from KG enhancement to guide the cross-view contrastive learning process, improve the representation learning ability and further suppress noise.
\end{itemize}

\subsubsection{Experimental Setup}
To evaluate the performance of the model, we adopt the commonly used evaluation criteria~\cite{yang2022knowledge, xia2022hypergraph} Recall@$K$ and NDCG@$K$, and set $K=20$.
For other baselines, we set the batch size to 1024, the learning rate of model optimization to $1e^{-3}$, and the embedding dimension to 64.
All training parameters are initialized by Xavier~\cite{glorot2010understanding}, and the network weights are updated iteratively by Adam optimizer~\cite{kingma2014adam}.
For SDK, we set the hidden state dimensionality to 32, the batch size to 1024, $\alpha = 0.5$ in the weighted sum function $\gamma$, and $K = 8$ in the hypergraph construction to select the relevant instances.
Grid search is adopted to confirm the optimal hyperparameters.
The learning rate is searched in $\{1e^{-3}, 5e^{-4}, 1e^{-4}, 1e^{-5}\}$, the $L_2$ regularization coefficient $\lambda_2$ in $\{1e^{-2}, 1e^{-3}, \cdots, 1e^{-6}\}$, the contrastive loss coefficient $\lambda_1$ in $\{2e^{-2},2e^{-3},\cdots,2e^{-6}\}$, the temperature $\tau$ in $\{0.1,0.25, 0.5, 0.75, 1.0 \}$, the hypergraph convolution layer $l$ in $\{ 1,2,3,4 \}$.
\subsection{Performance Analysis (RQ1)}

\begin{table}[htbp]
  \caption{Performance comparison with all methods on Yelp2018, Amazon-book and MIND.}
  \label{table2}
  \setlength{\tabcolsep}{0.9mm}{
    \begin{tabular}{c|cc|cc|cc}
      \hline
      \multirow{2}*{\diagbox{model}{dataset}} & \multicolumn{2}{c|}{Yelp2018} & \multicolumn{2}{c|}{Amazon-book} & \multicolumn{2}{c}{MIND}                                                       \\
                                              & Recall                        & NDCG                             & Recall                   & NDCG            & Recall          & NDCG            \\
      \hline
      BPR                                     & 0.0317                        & 0.0215                           & 0.0352                   & 0.0223          & 0.1072          & 0.0766          \\
      NCF                                     & 0.0389                        & 0.0256                           & 0.0408                   & 0.0251          & 0.1189          & 0.0831          \\
      KGAT                                    & 0.0469                        & 0.0316                           & 0.0503                   & 0.0289          & 0.1235          & 0.0876          \\
      LightGCN                                & 0.0532                        & 0.0378                           & 0.0556                   & 0.0318          & 0.1277          & 0.0951          \\
      MVIN                                    & 0.0551                        & 0.0403                           & 0.0572                   & 0.0346          & 0.1321          & 0.0986          \\
      CKAN                                    & 0.0547                        & 0.0396                           & 0.0589                   & 0.0350          & 0.1308          & 0.0973          \\
      DHCF                                    & 0.0504                        & 0.0363                           & 0.0527                   & 0.0296          & 0.1256          & 0.0927          \\
      HCCF                                    & 0.0598                        & 0.0423                           & 0.0588                   & 0.0358          & 0.1357          & 0.1018          \\
      SGL                                     & 0.0578                        & 0.0409                           & 0.0602                   & 0.0359          & 0.1328          & 0.0993          \\
      KGCL                                    & 0.0620                        & 0.0435                           & 0.0615                   & 0.0372          & 0.1380          & 0.1035          \\
      \hline
      \textbf{SDK}                            & \textbf{0.0663}               & \textbf{0.0467}                  & \textbf{0.0644}          & \textbf{0.0395} & \textbf{0.1427} & \textbf{0.1073} \\
      \hline
    \end{tabular}
  }
\end{table}

\noindent We conduct an overall performance evaluation on three datasets with all baselines to demonstrate the effectiveness of SDK. The results are shown in Table \ref{table2}.
\begin{itemize}
  \item SDK consistently outperforms all baselines on different datasets, which verifies the superiority and effectiveness of our model.
        Specifically, it achieves remarkable improvements over the highest baselines w.r.t. Recall@20 by 6.94\%, 4.72\%, and 3.41\% in Yelp2018, Amazon-book, and MIND, respectively.
        The advantages of SDK can be attributed to two aspects: (1) Benefiting from the HKG modeling, SDK can study the item-wise semantics of KG in $N$-ary contexts to obtain fine-grained item features.
        (2) SDK adopts the hypergraph self-supervised learning framework to distill the supervision signals from local and global perspectives in a cross-view way and then guide the model prediction process.
  \item From the above evaluation results, most KG-based methods (e.g., MVIN, CKAN) have achieved better performance than traditional CF methods (e.g., NCF, BPR).
        This confirms that the introduction of KG as side information can help alleviate the data sparsity problem of CF.
        However, the KG-based method KGAT is inferior to LightGCN, which is a basic GNN paradigm without KG enhancement.
        We argue that this gap is due to insufficient KG mining.
        The CF part cannot obtain quality representations from coarse-grained item features, but is disturbed by the unclear information of the KG to damage the model performance.
        SDK obtains fine-grained features by modeling the HKG, and dynamically constructs the hypergraph learning process based on this to obtain robust representations and alleviate over-smoothing problems.
  \item The relatively best performance of KGCL indicates the rationality of the KG-based method assisted by self-supervised learning.
        Moreover, HCCF follows the hypergraph self-supervised learning paradigm and achieves better performance than DHCF and SGL, showing that hypergraph structure can capture global dependency effectively.
        Self-supervised learning demands to generate supervision signals through view comparison to guide model training, and the semantic richness of KG ensures the robustness of the generated views.
        SDK dynamically constructs a hypergraph view from the refined item representations, and combines the local CF view to generate supervision signals in a cross-view manner to alleviate the problem of interaction sparsity.
\end{itemize}

\subsection{Ablation study (RQ2)}
\begin{table}[h!]
  \caption{Impact study with variants of SDK.}
  \label{table3}
  \centering
  \setlength{\tabcolsep}{0.9mm}{
    \begin{tabular}{c|cc|cc|cc}
      \hline
      \multirow{2}*{\diagbox{model}{dataset}} & \multicolumn{2}{c|}{Yelp2018} & \multicolumn{2}{c|}{Amazon-book} & \multicolumn{2}{c}{MIND}                                                       \\
                                              & Recall                        & NDCG                             & Recall                   & NDCG            & Recall          & NDCG            \\
      \hline
      \textbf{SDK}                            & \textbf{0.0663}               & \textbf{0.0467}                  & \textbf{0.0644}          & \textbf{0.0395} & \textbf{0.1427} & \textbf{0.1073} \\
      -SA                                     & 0.0635                        & 0.0445                           & 0.0614                   & 0.0378          & 0.1364          & 0.1036          \\
      -DH                                     & 0.0648                        & 0.0458                           & 0.0635                   & 0.0383          & 0.1383          & 0.1051          \\
      -SSL                                    & 0.0621                        & 0.0429                           & 0.0593                   & 0.0358          & 0.1376          & 0.1039          \\
      \hline
    \end{tabular}
  }
\end{table}

\noindent To get deep insights on our model, we generate several variants by simplifying key components to verify their positive contributions to recommendation performance.

\subsubsection{Investigation of interdependencies}
By removing the self-attention part of hyper-relational encoder, termed as \textit{-SA}, we investigate the effect of interdependencies between entities on the model.
From Table \ref{table3}, we observed that the simplification of the module resulted in a decrease in recommendation accuracy.
This phenomenon verifies the rationality of pre-capturing the interdependencies between entities before aggregating hyper-relational facts, and the simplified variant of \textit{-SA} increases the difficulty of effective aggregation.
\subsubsection{Investigation of self-supervision}
We also generate two variants to explore the effectiveness of components in dynamic hypergraph self-supervised learning.
Specifically, we establish a variant \textit{-DH} by prohibiting the hypergraph structure from changing dynamically with the number of layers, that is, only the initially constructed hypergraph is applied in the entire convolution process.
Another variant \textit{-SSL} is generated by removing the self-supervised learning task between global hypergraph dependencies and local collaborative views.
As shown in Table \ref{table3}, these variants have varying degrees of performance degradation, confirming the effectiveness of our dynamic hypergraph learning paradigm.
Since the initially constructed hypergraph may not be suitable for data representation, our model dynamically updates the global dependency to the hypergraph structure of the next layer to obtain richer node features.
Further, we follow the self-supervised learning paradigm to improve the representation learning ability by discriminating the same or different entities on the global and local views to obtain additional supervision signals.

\subsection{Benefits of SDK (RQ3)}
\subsubsection{Impact of smoothness degree}
\begin{table}[h!]
  \caption{Comparative evaluation of graph smoothness degree of variants (measured by MAD).}
  \label{table4}
  \centering
  \setlength{\tabcolsep}{0.9mm}{
    \begin{tabular}{c|cc|cc|cc}
      \hline
      \multirow{2}*{\diagbox{model}{dataset}} & \multicolumn{2}{c|}{Yelp2018} & \multicolumn{2}{c|}{Amazon-book} & \multicolumn{2}{c}{MIND}                                                       \\
                                              & User                          & Item                             & User                     & Item            & User            & Item            \\
      \hline
      \textbf{SDK}                            & \textbf{0.9436}               & \textbf{0.9315}                  & \textbf{0.9327}          & \textbf{0.9306} & \textbf{0.9766} & \textbf{0.9673} \\
      -SA                                     & 0.9317                        & 0.9125                           & 0.9152                   & 0.9102          & 0.9414          & 0.9347          \\
      -DH                                     & 0.8979                        & 0.8824                           & 0.8743                   & 0.8618          & 0.9255          & 0.9263          \\
      \hline
    \end{tabular}
  }
\end{table}

\noindent SDK enriches the item features by modeling the interdependencies, and the construction of the dynamic hypergraph also enables the refined features to be retained in the deep vector space.
Therefore, to validate whether SDK alleviates the over-smoothing problem, we compare the Mean Average Distance (MAD) between SDK and the two variants \textit{-SA} and \textit{-DH} to evaluate the effectiveness of the model.
As shown in Table \ref{table4}, we can observe that the MAD scores of the two variants are relatively low, indicating that their over-smoothing problem may be more serious.
In general, SDK can obtain a more qualitative representation by enriching the initial features and applying the dynamically adapted hypergraph structure to prevent feature convergence during the propagation process. While, the introduction of supervision signal improves the embedding quality and makes it less likely to cause smoothing.
\subsubsection{Impact of supervision signal sparsity}
\begin{figure}[htbp]
  \centering
  \subfigure[Yelp2018]{\includegraphics[width=4.2cm]{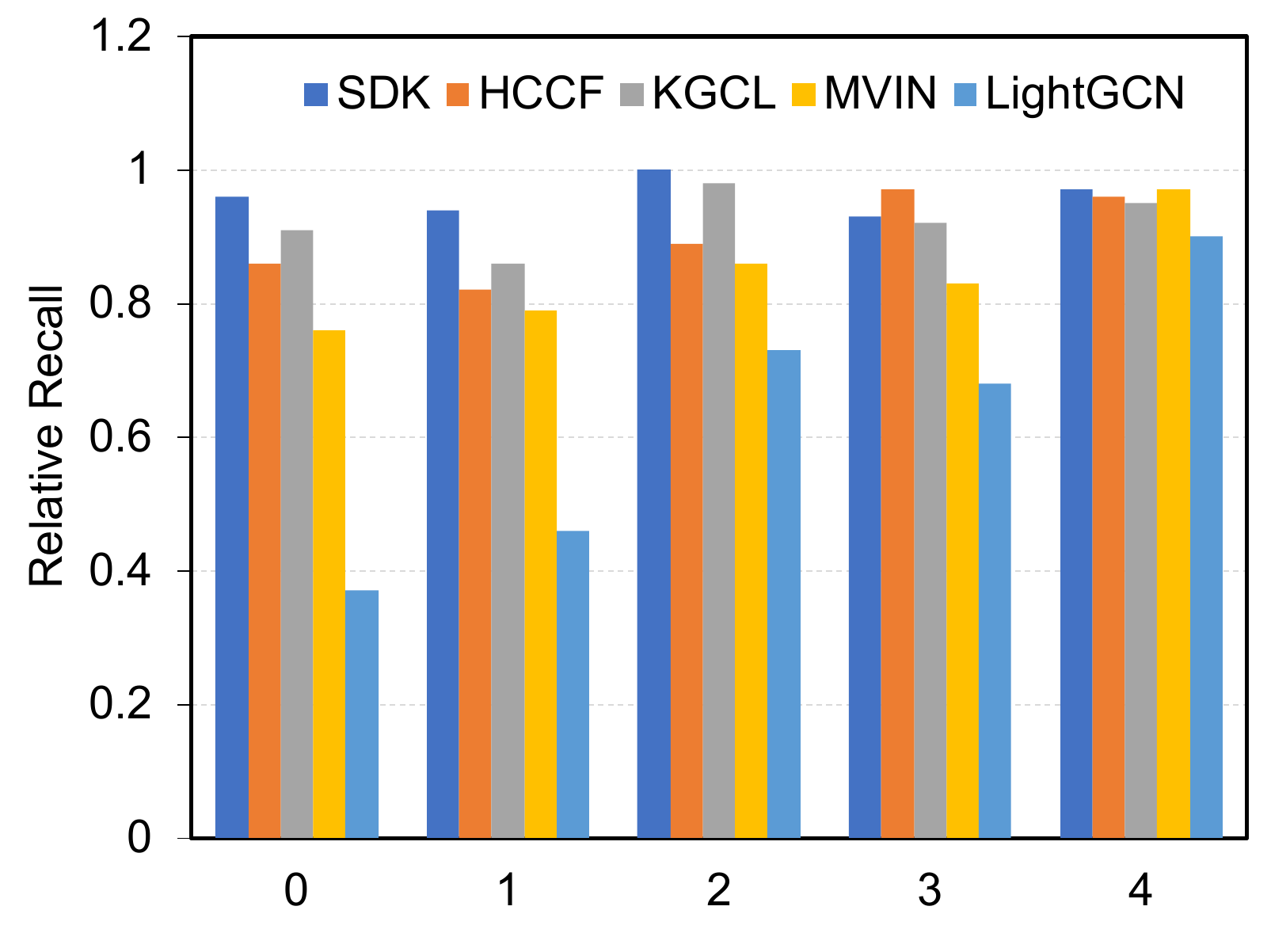}}
  \subfigure[MIND]{\includegraphics[width=4.2cm]{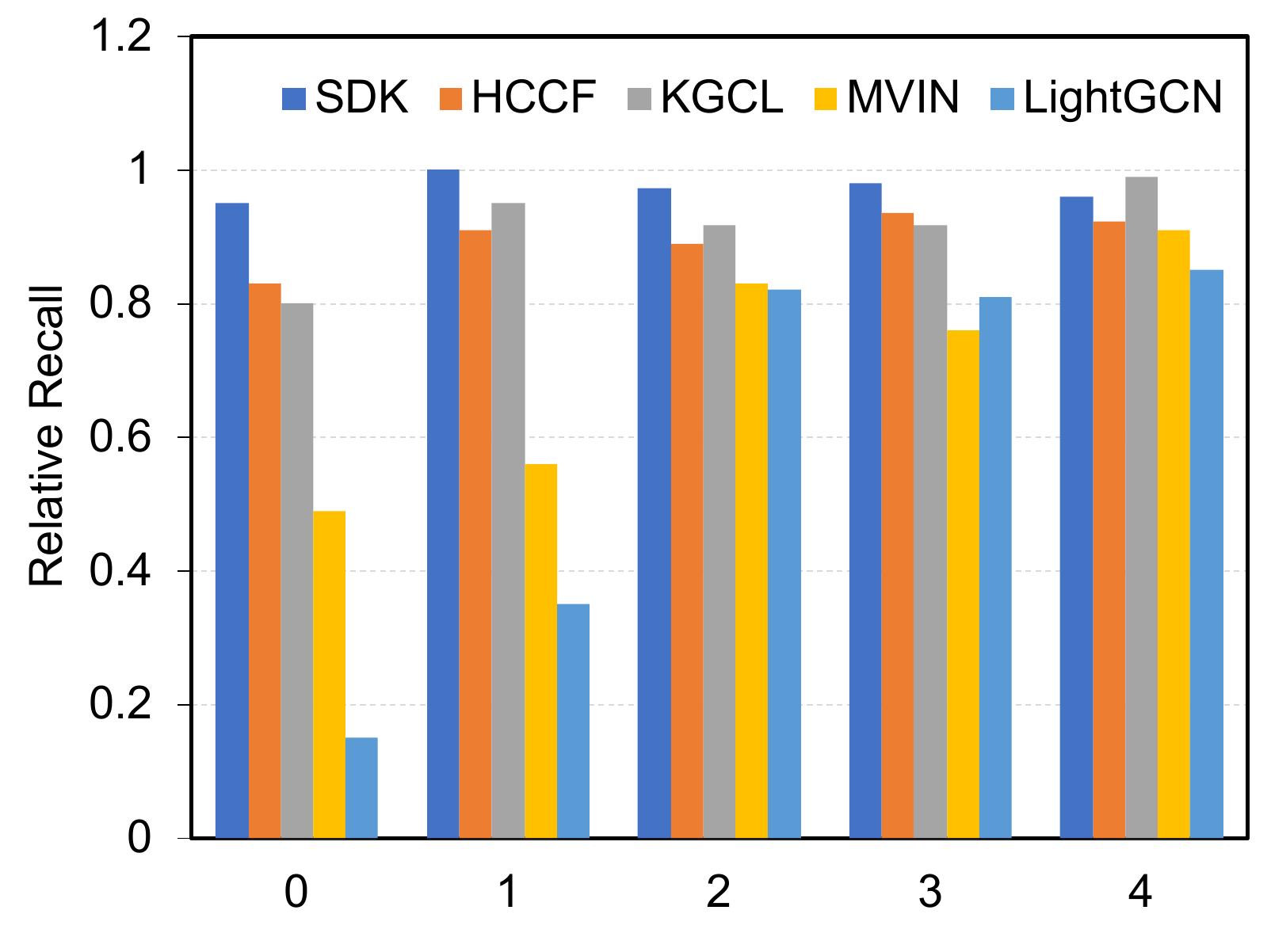}}
  \caption{Performance comparison under different interaction density degree groups.}
  \label{figsparse}
\end{figure}

\noindent Next, we group items according to the number of connections to verify whether SDK is robust enough to handle items without sufficient interactions.
The larger the group ID, the more connections the items have, which also means the greater the interaction density.
From Figure \ref{figsparse}, we observe that SDK performs best in most density degree groups.
In particular, in the group with extremely sparse supervision signals, the performance degradation is relatively gentle, which demonstrates that SDK is still robust in special scenarios.
Since SDK follows the self-supervised learning paradigm to distill the supervision signals from the global and local views, it compensates for the lack of supervision signals and effectively alleviates the popularity bias.

\subsection{Hyperparameter Analysis (RQ4)}
\begin{figure}[htbp]
  \centering
  \subfigure[Temperature]{
    \includegraphics[width=4.2cm]{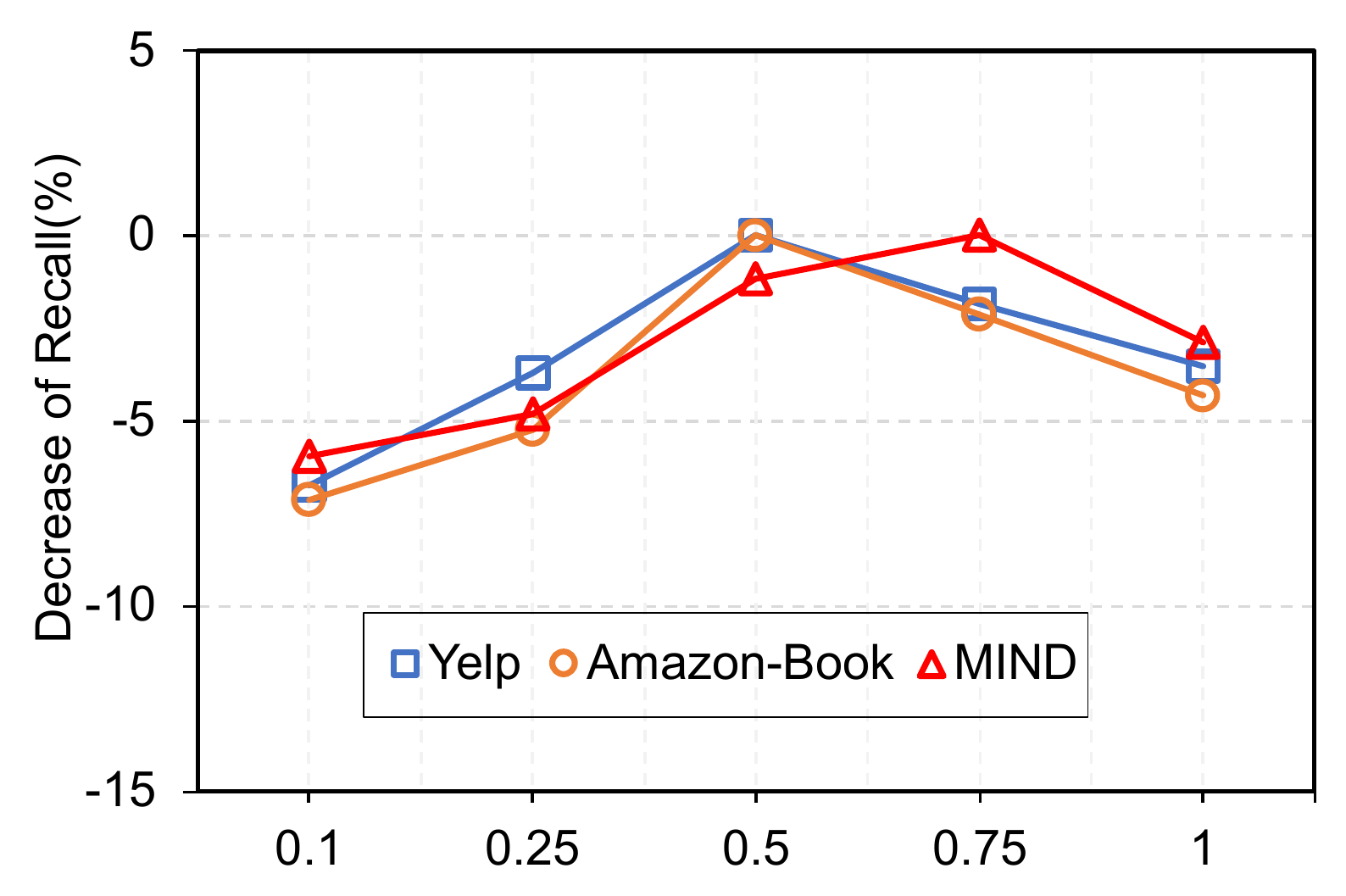}
    \includegraphics[width=4.2cm]{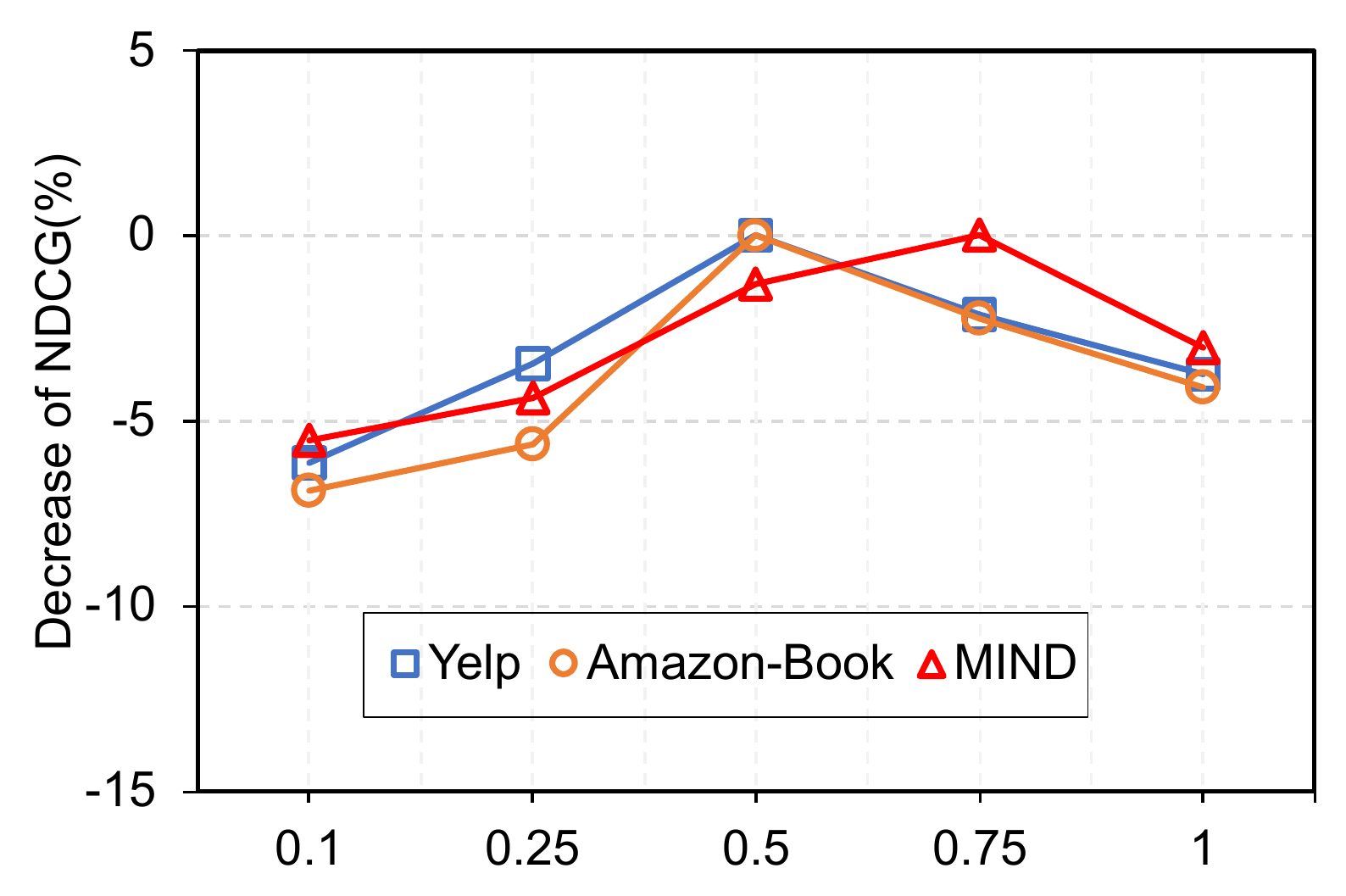}
  }
  \subfigure[Layer]{
    \includegraphics[width=4.2cm]{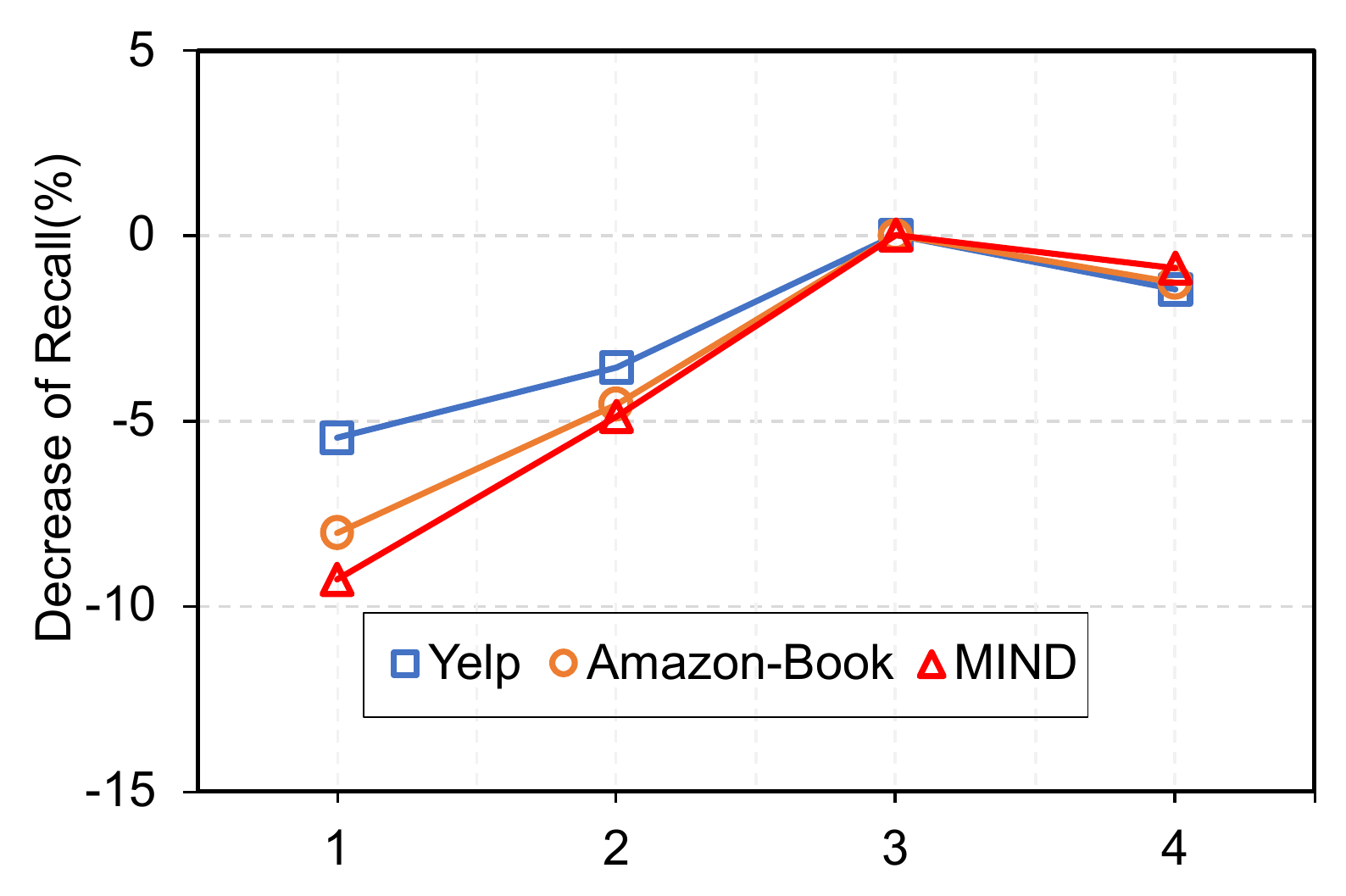}
    \includegraphics[width=4.2cm]{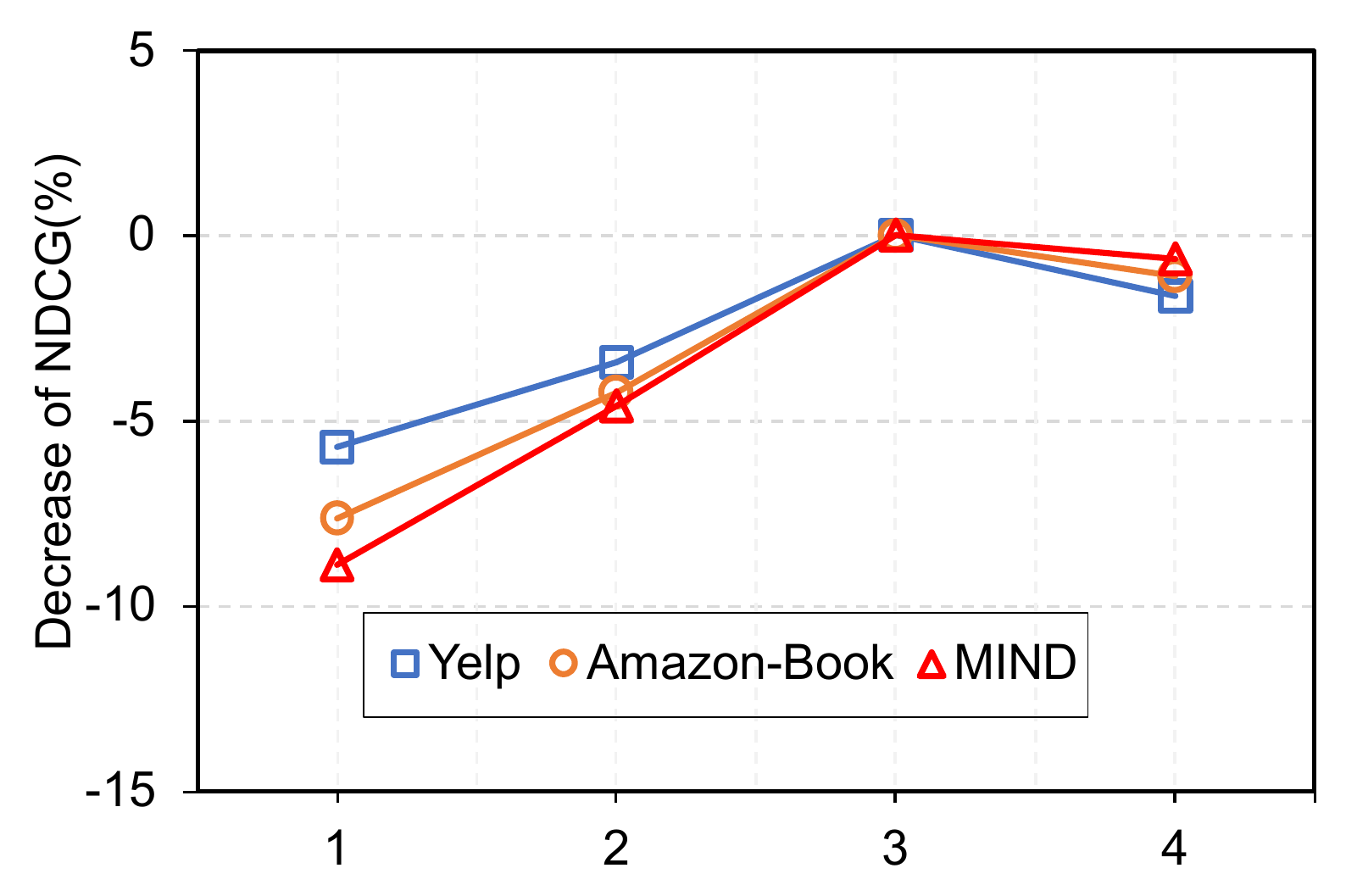}
  }
  \caption{Hyperparameter sensitivity analysis of temperature coefficient $\tau$ and hypergraph convolution layers $l$.}
  \label{figpara}
\end{figure}
\noindent In this section, we study the impact of the selection of key hyperparameters on the model, including the number of hypergraph convolution layers $K$ and the temperature parameter $\tau$, and report the evaluation results in Figure \ref{figpara}.
\begin{itemize}
  \item The temperature parameter $\tau$ is used to control the model's discrimination of negative samples.
        Improper temperature parameters may cause the model to pay too much attention to indistinguishable negative samples or lose bias attention to negative samples, resulting in performance degradation.
        From Figure \ref{figpara}, $\tau=0.5$ works best in Yelp2018 and Amazon-book, while $\tau=0.75$ is a more reasonable choice in MIND indicates that tolerating some difficult samples can achieve better performance.
  \item The deepening of the hypergraph convolution layer can expand the receptive field of neighborhood aggregation, but it may also face the problem of over-smoothing.
        In our model, rich semantic facts and dynamic hypergraph construction encourage the retention and propagation of features, thus maintaining good performance at a deeper level.
\end{itemize}

\section{Conclusion}
In this work, we propose SDK, a novel hypergraph self-supervised recommendation model based on HKG.
Specifically, we apply hyper-relational encoding to model the rich $N$-ary semantic knowledge of HKG to benefit representation learning.
In addition, the dynamic hypergraph convolution module in SDK effectively preserves the features into the deep vector space, alleviating the issue of over-smoothing.
SDK follows the self-supervised learning paradigm and generates supervision signals to guide the model learning process in a cross-view manner.
Extensive experiments on different datasets demonstrate the effectiveness of our model over other state-of-the-art methods.
For future work, we may explore how to solve the noise interference caused by KG enhancement.

\section{Acknowledgments}
This work is partly supported by National Natural Science Foundation of China (62172351), the "14th Five-Year Plan" Civil Aerospace Pre-Research Project of China (D020101), the Fund of Prospective Layout of Scientific Research for NUAA (Nanjing University of Aeronautics and Astronautics), Australian Research Council Future Fellowship (FT210100624), Discovery Project (DP190101985), Postgraduate Research \& Practice Innovation Program of NUAA (xcxjh20221605).
\printbibliography

\end{document}